\documentclass[12pt]{article}
\usepackage{axodraw,bbold}

\catcode`@=12
\begin{document}
\thispagestyle{empty}
\begin{flushright} 
UCRHEP-T410\\ 
May 2006\
\end{flushright}
\vspace{0.5in}
\begin{center}
{\LARGE	\bf Common Origin of Neutrino Mass,\\
Dark Matter, and Baryogenesis\\}
\vspace{1.5in}
{\bf Ernest Ma\\}
\vspace{0.2in}
{\sl Physics Department, University of California, Riverside, 
California 92521\\}
\vspace{1.5in}
\end{center}
\begin{abstract}\
Combining one established idea with two recent ones, it is pointed 
out for the first time that three of the outstanding problems of 
particle physics and cosmology, i.e. neutrino mass, dark matter, 
and baryogenesis, may have a simple common solution, arising from 
the interactions of a single term, with experimentally verifiable 
consequences.
\end{abstract}

\newpage
\baselineskip 24pt

The minimal standard model (SM) of particle interactions is very 
successful, but it fails to address three of the most important 
qusetions facing particle physics and cosmology today: (1) How is 
a neutrino mass generated?  (2) What is the nature of cold dark 
matter? (3) Why is there more matter than antimatter in the present 
Universe?  Each question has several different good answers, but 
they are generally unrelated.  In this note, combining one 
established idea and two recent ones, it is pointed out for the 
first time that all three may have a simple common origin, arising 
from the interactions of a single term, with experimentally verifiable 
consequences.

It has been known for a long time that neutrino mass and baryogenesis 
may be related through the canonical seesaw mechanism \cite{seesaw1,
seesaw2,seesaw3} and leptogenesis \cite{lg}.  The idea is very simple.  
The SM is augmented by three heavy neutral singlet fermions $N_i$
(often referred to as right-handed neutrinos) which have large 
Majorana masses $M_i$.  Through the Yukawa couplings
\begin{equation}
{\cal L}_Y = h_{\alpha i} (\nu_\alpha \phi^0 - l_\alpha \phi^+) N_i 
+ H.c.,
\end{equation}
where $(\nu_\alpha, l_\alpha)$, $\alpha = e,\mu,\tau$, are the three 
left-handed lepton doublets, and $(\phi^+,\phi^0)$ is the SM Higgs 
scalar doublet, the unique dimension-five operator for Majorana 
neutrino mass \cite{w79}
\begin{equation}
{\cal L}_\Lambda = {f_{\alpha \beta} \over 2\Lambda} (\nu_\alpha \phi^0 -
l_\alpha \phi^+) (\nu_\beta \phi^0 - l_\beta \phi^+) + H.c.
\end{equation}
is realized with $f_{\alpha \beta}/\Lambda = \sum_i h_{\alpha i} 
h_{\beta i}/M_i$.  As $\phi^0$ acquires a nonzero vacuum expectation 
value $\langle \phi^0 \rangle = v$, a neutrino mass matrix 
$({\cal M}_\nu)_{\alpha \beta} = f_{\alpha \beta} v^2/\Lambda$ is 
generated \cite{seesaw1,seesaw2,seesaw3}.  At the same time, the heavy 
Majorana fermions $N_i$ may decay into either $\nu \phi^0$ and $l^- \phi^+$ 
or their antiparticles in the early Universe, and establish a lepton asymmetry 
\cite{lg}, which gets converted into the present observed baryon 
asymmetry of the Universe through sphalerons at the electroweak 
phase transition \cite{krs}.  The interactions of Eq.~(1) together 
with $M_i$ are thus responsible for both neutrino mass and 
baryogenesis.  What is missing is just dark matter. 

To include dark matter, it has recently been recognized \cite{m06} 
that a very simple change of Eq.~(1) is all one needs, i.e.
\begin{equation}
{\cal L}_Y = h_{\alpha i} (\nu_\alpha \eta^0 - l_\alpha \eta^+) N_i 
+ H.c.,
\end{equation}
where $(\eta^+,\eta^0)$ is a second scalar doublet, and an exactly 
conserved $Z_2$ symmetry has been assumed, under which $(\eta^+,\eta^0)$ 
and $N_i$ are odd, with all other particles even.  This means that 
$\langle \eta^0 \rangle = 0$ and $N_i$ are \underline{not} the Dirac 
mass partners of $\nu_\alpha$ as in the canonical seesaw model. 
Nevertheless, a Majorana neutrino mass matrix ${\cal M}_\nu$ is 
obtained in one loop \cite{m98}, as depicted in FIG.~1.

\begin{figure}[htb]
\begin{center}
\begin{picture}(360,120)(0,0)
\ArrowLine(110,10)(150,10)
\ArrowLine(180,10)(150,10)
\ArrowLine(180,10)(210,10)
\ArrowLine(250,10)(210,10)
\ArrowLine(180,55)(150,10)
\ArrowLine(180,55)(210,10)
\ArrowLine(160,85)(180,55)
\ArrowLine(200,85)(180,55)

\Text(130,0)[]{$\nu_\alpha$}
\Text(230,0)[]{$\nu_\beta$}
\Text(180,0)[]{$N_i$}
\Text(155,40)[]{$\eta^{0}$}
\Text(210,40)[]{$\eta^{0}$}
\Text(158,95)[]{$\phi^{0}$}
\Text(210,95)[]{$\phi^{0}$}

\end{picture}
\end{center}
\caption{One-loop generation of neutrino mass.}
\end{figure}
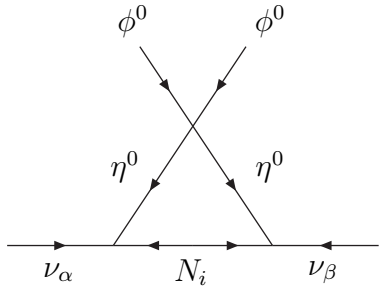

Let $\eta^0 = (\eta^0_R + i \eta^0_I)/\sqrt 2$ and $m_{R,I}$ be the 
mass of $\eta^0_{R,I}$, then
\begin{equation}
({\cal M}_\nu)_{\alpha \beta} = \sum_i {h_{\alpha i} h_{\beta i} M_i 
\over 16 \pi^{2}} \left[ {m_R^{2} \over m_R^{2}-M_i^{2}} \ln {m_R^{2} 
\over M_i^{2}} - {m_I^{2} \over m_I^{2}-M_i^{2}} \ln {m_I^{2} \over 
M_i^{2}} \right].
\end{equation}

The immediate consequence of the exact $Z_2$ symmetry of this model is 
that the lightest $N_i$ or $\eta^0_{R,I}$ is a candidate for the cold 
dark matter of the Universe.  Assuming that $N_i$ is the cold dark 
matter, it has been shown \cite{kms} that its relic abundance is 
compatible with observation only for $m_{R,I}$ less than about 350 GeV, 
for which the experimental upper bound of $1.2 \times 10^{-11}$ on the 
branching fraction of $\mu \to e \gamma$ implies a lower bound on the 
mass of the lightest neutrino of order 0.05 eV.  This result is of course 
very interesting, but it goes against leptogenesis because the 
interactions of Eq.~(3) must then be lepton-number violating at the 
electroweak scale and would wash out any lepton asymmetry even if it were 
somehow created at higher energies.

There is however another option.  Instead of $N_i$, consider $\eta^0_R$ 
or $\eta^0_I$ as the cold dark matter \cite{bhr}.  In that case, its 
relic abundance does not have to depend on the interactions of Eq.~(3), 
but rather on its gauge and scalar interactions.  This allows $M_i$ to 
be much heavier than $m_{R,I}$ in Eq.~(4) which reduces to
\begin{equation}
({\cal M}_\nu)_{\alpha \beta} = \sum_i {h_{\alpha i} h_{\beta i} 
\over 16 \pi^{2} M_i} \left[ m_R^{2} \ln {M_i^{2} \over m_R^{2}} - 
m_I^{2} \ln {M_i^{2} \over m_I^{2}} \right].
\end{equation}
Compared to the canonical seesaw formula of $({\cal M}_\nu)_{\alpha 
\beta} = \sum_i h_{\alpha i} h_{\beta i} v^2/M_i$, it is reduced  
by a factor of about $16 \pi^2$, assuming $M_R^2 \ln (M_i^2/M_R^2) - M_I^2 
\ln (M_i^2/M_I^2)$ is of order $v^2$.  This means that the Davidson-Ibarra 
bound \cite{di} on leptogenesis is now reduced by $16 \pi^2$, and 
the lightest $N_i$ needs only to be heavier than about $2.5 \times 10^7$ GeV, 
well below a possible gravitino bound of $10^9$ GeV. In the decay of $N_i$ 
to either $\nu \eta^0$ and $l^- \eta^+$ or their antiparticles, not 
only is a lepton asymmetry created, but also dark matter.  At 
temperatures below the smallest $M_i$, of order $10^7$ to $10^{11}$ 
GeV, the interactions of Eq.~(3) are frozen out and the lepton 
asymmetry remains and will be converted to the observed baryon 
asymmetry of the Universe through sphalerons in the usual way \cite{lg_rev}.  
At the same time, because of the $\lambda_5$ term to be discussed later, 
there is no corresponding dark-matter asymmetry, and the annihilation of 
$\eta^0$ at a temperature when it becomes nonrelativistic will 
determine its relic abundance \cite{dm_rev}, as shown in Ref.~\cite{bhr}. 
The stated goal of having a simple common origin, i.e. Eq.~(3), for 
neutrino mass, dark matter, and baryogenesis has been realized.

The next question to consider is this model's experimental consequences. 
At the electroweak scale, its particle content is identical to that of 
Ref.~\cite{bhr}, except that neutrinos have mass here according to Eq.~(5), 
but not there.  The scalar sector consists of two doublets: 
$(\phi^+,\phi^0)$ as in the SM, and the new all-important $(\eta^+,\eta^0)$.  
Their scalar interactions are given by
\begin{eqnarray}
V &=& m_1^{2} \Phi^\dagger \Phi + m_2^{2} \eta^\dagger \eta + {1 \over 2} 
\lambda_1 (\Phi^\dagger \Phi)^{2} + {1 \over 2} \lambda_2 
(\eta^\dagger \eta)^{2} + \lambda_3 (\Phi^\dagger \Phi)(\eta^\dagger \eta) 
\nonumber \\ &+& \lambda_4 (\Phi^\dagger \eta)(\eta^\dagger \Phi) + 
{1 \over 2} \lambda_5 [(\Phi^\dagger \eta)^{2} + H.c.],
\end{eqnarray}
where $\lambda_5$ has been chosen real without any loss of generality. 
For $m_1^{2} < 0$ and $m_2^{2} > 0$, only $\phi^{0}$ acquires a nonzero 
vacuum expectation value $v$.  The masses of the resulting physical scalar 
bosons are given by
\begin{eqnarray}
m^{2} (\phi^{0}_R) &=& 2 \lambda_1 v^{2}, \\ 
m^{2} (\eta^{\pm}) &=& m_2^{2} + \lambda_3 v^{2}, \\ 
m^{2} (\eta^{0}_R) &=& m_2^{2} + (\lambda_3 + \lambda_4 + 
\lambda_5) v^{2}, \\ 
m^{2} (\eta^{0}_I) &=& m_2^{2} + (\lambda_3 + \lambda_4 - 
\lambda_5) v^{2}.
\end{eqnarray}
Since $\eta^{\pm}$ cannot be dark matter, $\lambda_4 \pm \lambda_5 < 0$ 
is required.  If $\lambda_5 < 0$ is also chosen, then $\eta^0_R$ is 
lighter than $\eta^0_I$, and becomes the dark-matter candidate of this 
model.

The $\eta$ particles can be produced in pairs directly by the SM gauge bosons 
$W^{\pm}$, $Z$, or $\gamma$.  Once produced, $\eta^{\pm}$ will decay into 
$\eta^0_{R,I}$ and a virtual $W^{\pm}$, which becomes a quark-antiquark 
or lepton-antilepton pair, and $\eta^0_I$ will decay into $\eta^0_R$ and 
a virtual $Z^0$.  The decay chain
\begin{equation}
\eta^+ \to \eta^0_I l^+ \nu, ~~~{\rm then} ~~\eta^0_I \to \eta^0_R l^+ l^-
\end{equation}
has 3 charged leptons and large missing energy, and can be compared to 
the direct decay
\begin{equation}
\eta^+ \to \eta^0_R l^+ \nu
\end{equation}
to extract the masses of the respective particles.

The direct detection of $\eta^0_R$ in present dark-matter search 
experiments is discussed in Ref.~\cite{bhr}, where it 
is found to be about two orders of magnitude below 
current experimental capabilities.  The additional 
contribution from Eq.~(3) is negligible because $\eta^0_R$ interacts 
directly only with neutrinos, and $N_i$ are assumed very heavy. 
Contribution to the one-loop radiative process $\mu \to e \gamma$ 
is also negligible, in contrast to the case where $N_i$ play the role 
of dark matter \cite{kms}.

To conclude, it has been pointed out in this note that neutrino mass, 
dark matter, and baryogenesis may have a simple common origin, i.e. 
Eq.~(3).  The proposed model \cite{m06} uses the result of Ref.~\cite{bhr} 
that $\eta^0_R$ is a suitable bosonic dark-matter candidate, and that 
of canonical leptogenesis \cite{lg}.  It predicts rather small cross 
sections for $\eta^0_R$ detection at low energies, although future 
experiments will become sensitive to it.  The decays of $\eta^{\pm}$ 
and $\eta^0_I$ into $\eta^0_R$ should be observable at the forthcoming 
Large Hadron Collider (LHC).

I thank M. Frigerio, T. Hambye, and J. Kubo for useful discussions. 
This work was supported in part by the U.~S.~Department of Energy 
under Grant No.~DE-FG03-94ER40837.

\newpage
\bibliographystyle{unsrt}

\end{document}